\pdfoutput=1               
\documentclass[aps,prb,twoside,twocolumn,10pt,superscriptaddress,longbibliography,nofootinbib,nobibnotes]{revtex4-2}

\usepackage[T1]{fontenc}
\usepackage[utf8]{inputenc}
\usepackage{lmodern}             
\usepackage{amsmath,amssymb}
\usepackage{physics}             
\usepackage{siunitx}             
\usepackage{graphicx}
\usepackage{microtype}           
\usepackage{xcolor}
\usepackage{booktabs}            
\usepackage{hyperref}
\hypersetup{colorlinks=true,linkcolor=blue,citecolor=blue,urlcolor=blue}
\usepackage[bottom]{footmisc}
\usepackage{comment}
\usepackage{adjustbox}
\usepackage{tablefootnote}
\usepackage{threeparttable}
\usepackage{titlesec}

\makeatletter
\@ifundefined{frontmatter@abstract@produce}{%
  \ClassWarning{revtex}{\string\frontmatter@abstract@produce\space not defined; cannot insert space}%
}{%
  \let\oldfrontmatter@abstract@produce\frontmatter@abstract@produce
  \def\frontmatter@abstract@produce{%
    \vskip 2.0em 
    \oldfrontmatter@abstract@produce
    \vskip 5.0em 
  }%
}
\makeatother

\expandafter\def\expandafter\normalsize\expandafter{%
    \normalsize%
    \setlength\abovedisplayskip{4pt}%
    \setlength\belowdisplayskip{10pt}%
    \setlength\abovedisplayshortskip{4pt}%
    \setlength\belowdisplayshortskip{8pt}%
}

\begin{document}

\title{AMaRaNTA: Automated First-Principles Exchange Parameters In 2D Magnets}

\author{Federico Orlando}
\affiliation{Physics Department - Politecnico di Milano, p.za Leonardo da Vinci 32, 20133 Milan, Italy}
\affiliation{Consiglio Nazionale delle Ricerche CNR-SPIN, c/o Università degli Studi “G. D’Annunzio”, 66100 Chieti, Italy}

\author{Andrea Droghetti}
\affiliation{Department of Molecular Sciences and Nanosystems, Ca’ Foscari University of Venice, via Torino 155, 30170 Mestre-Venice, Italy}
\affiliation{Consiglio Nazionale delle Ricerche CNR-SPIN, c/o Università degli Studi “G. D’Annunzio”, 66100 Chieti, Italy}

\author{Lorenzo Varrassi}
\affiliation{Dipartimento di Fisica e Astronomia, Università di Bologna, 40127 Bologna, Italy}

\author{Giuseppe Cuono}
\affiliation{Consiglio Nazionale delle Ricerche CNR-SPIN, c/o Università degli Studi “G. D’Annunzio”, 66100 Chieti, Italy}

\author{Cesare Franchini}
\affiliation{University of Vienna, Faculty of Physics and Center for Computational Materials Science, Kolingasse 14-16, Vienna, Austria}
\affiliation{Dipartimento di Fisica e Astronomia, Università di Bologna, 40127 Bologna, Italy}

\author{Paolo Barone}
\affiliation{Consiglio Nazionale delle Ricerche CNR-SPIN, Area della Ricerca di Tor Vergata, 00133 Rome, Italy}

\author{Antimo Marrazzo}
\affiliation{Scuola Internazionale Superiore di Studi Avanzati (SISSA), 34136 Trieste, Italy}

\author{Marco Gibertini}
\affiliation{Dipartimento di Scienze Fisiche, Informatiche e Matematiche, Università di Modena e Reggio Emilia, Via Campi 213/a, 41125 Modena, Italy}
\affiliation{Centro S3, Istituto Nanoscienze-CNR, Via Campi 213/a, 41125 Modena, Italy}

\author{Srdjan Stavrić}
\affiliation{Vinča Institute of Nuclear Sciences - National Institute of the Republic of Serbia, University of Belgrade, P. O. Box 522, RS-11001 Belgrade, Serbia}
\affiliation{Consiglio Nazionale delle Ricerche CNR-SPIN, c/o Università degli Studi “G. D’Annunzio”, 66100 Chieti, Italy}

\author{Silvia Picozzi}
\affiliation{Department of Materials Science, University of Milan – Bicocca, 20125 Milan, Italy}
\affiliation{Consiglio Nazionale delle Ricerche CNR-SPIN, c/o Università degli Studi “G. D’Annunzio”, 66100 Chieti, Italy}

\date{\today}

\vspace{1em}
\begin{abstract}
Two-dimensional (2D) magnets host a wide range of exotic magnetic textures, whose low-energy excitations and finite-temperature properties are typically described by effective spin models based on Heisenberg-like Hamiltonians. A key challenge in this framework is the reliable determination, from {\it ab initio} calculations, of exchange parameters and their anisotropic components, crucial for stabilising long-range order. Among the different strategies proposed for this task, the energy-mapping method -- based on total-energy calculations within Density Functional Theory (DFT) -- is the most widely adopted, but it typically requires laborious, multi-step procedures. To overcome this limitation, we introduce AMaRaNTA (Automating Magnetic paRAmeters iN a Tensorial Approach), a computational package that systematically automates the energy-mapping method, specifically through its ``four-state'' formulation, to extract exchange and anisotropy parameters in 2D magnets. In its current implementation, AMaRaNTA returns the nearest-neighbour exchange tensor, complemented by scalar parameters for second- and third-nearest-neighbour exchange interactions as well as single-ion anisotropy. Together, these provide a minimal yet sufficient set of parameters to capture magnetic frustration and anisotropies, essential for stabilising several observed magnetic states in 2D materials. Applied to a representative subset of the Materials Cloud 2D Structure database, AMaRaNTA demonstrates robust, automated and reproducible screening of magnetic interactions, with clear potential for high-throughput simulations.
\end{abstract}

\maketitle


\setlength{\parskip}{0.1em}  
\section*{INTRODUCTION}

Magnetism in layered van der Waals (vdW)-bonded systems has been known for decades \cite{Tsubokawa, Dillon}. However, a renewed surge of interest in these materials emerged between 2016 and 2018, following the experimental observation of intrinsic ferromagnetic (FM) order in CrI$_3$ down to the monolayer limit \cite{CrI3_first} and in few-layer CrGeTe$_3$ \cite{CrGeTe_first}, as well as antiferromagnetic (AFM) order in monolayer FePS$_3$ \cite{Wang_2016, Lee2016}. These findings demonstrated that stable magnetic order can persist in the two-dimensional (2D) limit.
Since then, several other materials have been reported to retain magnetic order upon exfoliation from their bulk vdW-bonded counterparts, and the library of 2D magnets continues to grow steadily \cite{Gong_2019, Gibertini2019,Jiang2021, genome_2022}. These systems have already been incorporated into prototype spintronic devices (e.g., \cite{CrI3-MTJ1,hbn2,Zhang2024}), with ongoing experimental efforts aiming at raising magnetic ordering temperatures \cite{FGT3,first_FeGaT_paper} and enhancing perpendicular magnetic anisotropy \cite{nanolett.aniso.mono.Ga} for improved device performance.

From a theoretical standpoint, the existence of magnetism in 2D systems is particularly intriguing. The celebrated Mermin-Wagner theorem \cite{MW} implies that long-range magnetic order is forbidden in 2D at any finite temperature, being completely destroyed by thermal spin fluctuations. However, this result holds strictly for systems with isotropic, short-range magnetic interactions, as described by the Heisenberg model. In real materials, uniaxial magnetic anisotropies may break the continuous spin-rotational symmetry, opening a gap in the spin-wave spectrum. This in turn suppresses the destabilising thermal fluctuations, allowing for finite-temperature magnetic order \cite{PhysRevB.60.1082,Lado_2017}.

Another reason for the growing interest in 2D magnets lies in their rich variety of magnetic phases, which extend well beyond conventional ferromagnetism and antiferromagnetism. For example, a recent theoretical work predicts altermagnetism in RuF$_4$ monolayers and other 2D compounds \cite{Sodequist2024May}. In these centrosymmetric systems, there is nonetheless an antisymmetric Dzyaloshinskii–Moriya interaction (DMI) \cite{DM1,DM2}, driven by spin–orbit coupling (SOC), between pairs of atoms that locally break space-inversion symmetry, inducing spin canting and weak ferromagnetism \cite{Milivojevic2024May}. Meanwhile, in systems where space-inversion symmetry is broken not only locally but also globally, the DMI \cite{DM1,DM2} acts as a chiral interaction, yielding well-defined chiral noncollinear and noncoplanar spin textures \cite{Rossler2006, PhysRevB.80.054416, Heinze2011}. Skyrmions, for instance, have been predicted in several so-called Janus 2D materials \cite{PhysRevB.109.094431, PhysRevB.102.241107, PhysRevB.101.060404,Ga2022,Cui}, such as MnSeTe and MnSTe \cite{Liang20}. Conversely, in centrosymmetric lattices with geometric frustration — such as triangular or Kagome lattices — AFM exchange interactions can give rise to highly degenerate ground states and complex magnetic behaviour, including the emergence of spin-spiral structures \cite{ PhysRevB.58.12049,Ramirez1994,Balents2010, cryst7050121}. Similar frustration effects may also emerge, even in other lattice geometries, in the presence of competing exchange interactions, for example between nearest-neighbour and second- or third-nearest-neighbour spins \cite{RASTELLI1979,SCHMIDT2017}. Notably, frustrated centrosymmetric systems can still host skyrmion-like spin textures even in the absence of DMI \cite{chakrabartty2025}. For example, a spontaneous antibiskyrmion lattice with unique topology, driven by the anisotropic 
symmetric exchange interaction, has been predicted in the semiconducting NiI$_2$ monolayer \cite{Amoroso2020}. Interestingly, the onset of helimagnetic phases can spontaneously break the inversion symmetry of the underlying crystal lattice, inducing a polar axis and enabling ferroelectricity \cite{Cheong2007,Tokura2010}, as demonstrated again in monolayer NiI$_2$ \cite{Fumega_2022,Song2022}.

Given these premises, accurate computational approaches for predicting the magnetic properties of 2D materials from first-principles are in high demand. Among them, Density Functional Theory (DFT) \cite{Kohn_nobel} has emerged as the preferred method, offering a relatively good balance between accuracy, computational efficiency and ease of implementation. 
Furthermore, various techniques and computational tools have been developed within DFT to predict magnetic properties, particularly to calculate exchange parameters and magnetic anisotropies \cite{PhysRevB.99.224412,PhysRevB.91.224405,TB2J,PhysRevB.108.214418,PhysRevB.99.214426}. The most established ones include the energy-mapping method \cite{Noodleman},
the torque approach, also known as the Liechtenstein–Katsnelson–Antropov–Gubanov (LKAG) method \cite{LIECHTENSTEIN198765, RevModPhys.95.035004}, based on the magnetic force theorem, and the spin-spiral method, which relies on the generalised Bloch theorem \cite{Sandratskii_1986, Sandratskii_1991} — each with its own advantages and limitations, depending on the specific system and magnetic phenomena under investigation. When integrated into high-throughput computational frameworks \cite{MC2D_1,Haastrup2018Sep,Torelli}, these techniques enable the rapid screening of candidate materials with targeted magnetic properties, thereby accelerating materials discovery and providing valuable guidance and support for experiments. 

For example, Torelli {\it et al.} \cite{Torelli} studied around 150 2D compounds, encompassed in the C2DB database \cite{Haastrup2018Sep,Gjerding2021Jul}. In their work, the DFT total energy differences between collinear FM and AFM configurations are mapped onto a spin model that includes nearest-neighbour isotropic Heisenberg exchange, on-site (single-ion) anisotropy and an approximate two-ion anisotropy.  
This study not only confirmed the magnetic ground states of well-known compounds, but also predicted the properties of several less-studied materials.
Despite these successes, the model employed in Ref. \cite{Torelli} shows some limitations and is often insufficient to capture the rich variety of possible 2D magnetic phases. 
Therefore, the same group later extended their study to an even larger dataset of 192 materials, whose magnetic texture was assessed exploring the possibilities for spin-spiral orders \cite{Sodequist_2023} using the generalised Bloch theorem \cite{Sandratskii_1986, Sandratskii_1991}. Nonetheless, the method is limited to systems that can be well described by a single-$\mathbf{q}$ spiral (with $\mathbf{q}$ being the spiral wave-vector) and does not capture complex multi-$\mathbf{q}$ configurations.
Furthermore, no systematic fitting of the microscopic magnetic parameters was performed therein.

To further extend the systematic application of DFT for predicting the magnetic properties of 2D materials, a valuable approach is represented by the so-called ``four-state'' method \cite{Xiang13, Li21,PhysRevB.84.224429}.
Within this framework, each magnetic parameter of a spin model is extracted from DFT total energies computed for four distinct relative orientations of the magnetic moments of two selected atoms, while all other magnetic moments are held fixed. This provides a direct estimate of the local interactions between atomic pairs. The method offers a clear advantage over conventional energy-mapping approaches, which typically rely on extensive global sampling of numerous magnetic configurations spanning the entire simulation cell. When implemented for noncollinear DFT calculations including SOC, the four-state method gives access to the fully tensorial -- hence anisotropic -- nature of the magnetic interactions, enabling their mapping onto general spin models capable of describing complex magnetic states \cite{Amoroso2020, Xiang13, Li21,PhysRevB.84.224429, Xiang_CrI3,PhysRevB.101.060404,PhysRevB.102.241107, PhysRevLett.124.087205}. It also naturally captures local interactions leading to canting effects between spin pairs that are difficult to predict with global sampling \cite{Milivojevic2024May}. However, despite these strengths, applying the four-state method in high-throughput calculations poses several challenges. These include the need to automatically select relevant spin configurations, construct suitable (and potentially large) supercells to capture long-range interactions, and perform a substantial amount of DFT calculations, equal to four times the number of spin model parameters to be determined \cite{Xiang13, Li21}. 

To address these challenges, we introduce AMaRaNTA (Automating MAgnetic paRAmeters iN a Tensorial Approach), a computational package designed to streamline the four-state approach for automated, scalable calculations.
AMaRaNTA is implemented as a workflow within AiiDA (Automated Interactive Infrastructure and Database for Computational Science) \cite{Aiida}, a robust platform for managing complex computational workflows with full data provenance and automation capabilities. The workflow interfaces with the Vienna Ab-initio Simulation Package (VASP) \cite{VASP1,Vasp2} (though it can be in principle adapted to other electronic structure codes) to perform accurate noncollinear DFT total energy calculations including SOC. Designed for user-friendliness, AMaRaNTA requires only a structure file as input; it then automates the setup of simulation cells, the submission of calculations to (and retrieval of results from) high-performance computing facilities through AiiDA and, ultimately, the evaluation of the exchange parameters. Specifically, AMaRaNTA computes the full exchange tensor for nearest-neighbour interactions and scalar isotropic exchange parameters for second- and third-nearest-neighbour ones, as well as single-ion anisotropy terms. These provide a minimal starting set to capture magnetic frustration and anisotropies underpinning the stabilisation of diverse magnetic states in 2D materials. However, the framework lends itself to extensions beyond the present implementation, for instance, to include longer-range interactions -- along with their fully tensorial character -- allowing mapping onto more complex spin models.

To illustrate its performance, AMaRaNTA is applied to a compact dataset of about thirty 2D magnetic compounds from the Materials Cloud 2D Structure database~\cite{MC2D_1,MC2D_2}, demonstrating robust and automated screening of magnetic interactions. We successfully recover the expected magnetic features for a number of previously characterised 2D magnets. Furthermore, for previously unreported systems, our calculations predict AFM isotropic exchange in NiF$_4$Tl$_2$, anisotropic symmetric exchange in MnBi$_2$Te$_4$, and antisymmetric exchange in VF$_4$ and VAgP$_2$Se$_6$.


\section*{RESULTS}

\vspace{0.25cm}
\subsection*{Four-state method\label{sec:ExpCompDetails}}
\vspace{0.2cm}
The four-state method \cite{Xiang13, Li21} maps the results of first-principles total energy calculations (here performed within DFT) onto a classical, Heisenberg-like model Hamiltonian, as anticipated in the Introduction. In the following, we outline the specific formulation and conventions adopted in AMaRaNTA, along with the relevant definitions.

\subsubsection*{Spin model Hamiltonian}\label{sec:spin model}

The AMaRaNTA effective spin Hamiltonian is based on the bilinear Heisenberg model and is expressed as

\begin{equation}\label{eq.spinH}
\begin{split}
    H = \sum_{\substack{i<j \\ (i,j) \in 1\text{NN}}}\mathbf{S_i}\cdot \mathbf{\underline{\underline{J}}^{(1)}_{ij}}\cdot\mathbf{S_j} + \sum_{\substack{i<j \\ (i,j) \in 2\text{NN}}}J^{(2)}_{ij} \mathbf{S_i}\cdot  \mathbf{S_j} 
    \\
    + \sum_{\substack{i<j \\ (i,j) \in 3\text{NN}}}J^{(3)}_{ij} \mathbf{S_i} \cdot \mathbf{S_j} + \sum_{i}A_{i} S_{iz}^2
\end{split}
\end{equation}
where $n\text{NN}$ (with $n=1,2,3$) denotes the set of $n$th nearest-neighbour atomic pairs, and

\begin{itemize}
    \item $\mathbf{\underline{\underline{J}}^{(1)}_{ij}}$ is the full nearest-neighbour second-rank exchange tensor. Each of its elements $J_{ij,\alpha\beta}^{(1)}$ describes the coupling between the spin component $S^\alpha_i$ of atom $i$ and the spin component $S^\beta_j$ of atom $j$, where $\alpha, \beta \in \{x, y, z\}$. 
    $\mathbf{\underline{\underline{J}}^{(1)}_{ij}}$ is commonly decomposed as \cite{Li21}

    \vspace{-1.5em}
    \begin{equation}    \label{eq.J1decomp}\mathbf{\underline{\underline{J}}^{(1)}_{ij}} = J^{(1)}_{ij}\mathbf{\underline{\underline{I}}}+\mathbf{\underline{\underline{K}}_{ij}}+\mathbf{\underline{\underline{D}}_{ij}}
    \end{equation}
    
    that is, into a Heisenberg isotropic exchange parameter $J^{(1)}_{ij} = \frac{1}{3}Tr\mathbf{\underline{\underline{J}}^{(1)}_{ij}}$, a traceless symmetric matrix $\mathbf{\underline{\underline{K}}_{ij}} = (\mathbf{\underline{\underline{J}}^{(1)}_{ij}}+\mathbf{\underline{\underline{J}}^{(1)T}_{ij}})/2 - J^{(1)}_{ij}\mathbf{\underline{\underline{I}}}  $ (giving rise to anisotropic exchange interactions of the Kitaev type \cite{Kitaev1, Kitaev2}) and an antisymmetric part $\mathbf{\underline{\underline{D}}_{ij}} = (\mathbf{\underline{\underline{J}}^{(1)}_{ij}}-\mathbf{\underline{\underline{J}}^{(1)T}_{ij}})/2$, whose three independent components form the Dzyaloshinskii–Moriya vector $\mathbf{{D}}$ \cite{DM1, DM2}.
    
    Thus, in our model, the two-ion anisotropy is treated more rigorously than in previous high-throughput studies (e.g., \cite{Torelli}, where it was included through a term of the form $H_{ij}^{ani} = \lambda\sum_{ij}S^z_{i}S^z_{j}$ in the spin Hamiltonian).
    
    \item $J^{(2)}_{ij}$ and $J^{(3)}_{ij}$ are the second- and third-nearest-neighbour scalar exchange parameters, respectively.
    
    \item $A_{i}$ is the single-ion anisotropy (SIA). In general, the SIA can be represented as a second-rank symmetric tensor, analogous to $\mathbf{\underline{\underline{J}}^{(1)}_{ij}}$. However, this fully tensorial description is often unnecessary for practical purposes. In many 2D systems, the SIA can be effectively characterised by the simplified scalar form adopted here, which captures easy-axis or easy-plane anisotropy (see also Ref. \cite{Xiang13}).
\end{itemize}

In total, for a chosen atomic site $i$, the model contains twelve parameters: the nine components of the nearest-neighbour exchange tensor, the scalar exchange parameters for the second- and third-nearest-neighbours, and the SIA.

According to our sign conventions, a negative (positive) exchange parameter describes FM (AFM) spin alignment. Similarly, positive values of the SIA correspond to easy-plane anisotropy and negative values to easy-axis anisotropy.

\subsubsection*{Calculation of the spin Hamiltonian parameters}
In this work, we restrict to materials containing a single magnetic species, where all magnetic atoms are crystallographically equivalent and thus possess the same spin magnitude $S$. The DFT calculations underlying the four-state method employ supercells large enough to both capture interactions up to the third-nearest-neighbours -- consistently with the spin model -- and to suppress spurious interactions between an atom and its periodic replicas.

To extract the parameters of the spin model Hamiltonian, the following different sets of four-state calculations are performed.

\vspace{0.35cm}
\paragraph*{Nearest-neighbour exchange tensor}
The calculations for each element $J_{ij,\alpha\beta}^{(1)}$ of the nearest-neighbour exchange tensor for a pair of magnetic atoms $(i,j)$ are performed within the noncollinear DFT framework with SOC included. We compute the energies $E_{ij,+\alpha,+\beta}$, $E_{ij,+\alpha,-\beta}$, $E_{ij,-\alpha,+\beta}$, and $E_{ij,-\alpha,-\beta}$, corresponding to the four configurations in which the spins of atom $i$ and $j$ are constrained along $\pm\hat{\alpha}$ and $\pm\hat{\beta}$ Cartesian directions, respectively, while the spins of all other magnetic atoms are fixed orthogonal to both $\alpha$ and $\beta$, forming a reference magnetic background. The desired parameter is then given by

\begin{equation}
\label{eq:j_tensor}
J_{ij,\alpha\beta}^{(1)} = \frac{
E_{ij,+\alpha,+\beta}
- E_{ij,+\alpha,-\beta}
- E_{ij,-\alpha,+\beta}
+ E_{ij,-\alpha,-\beta}
}{4S^2}
\end{equation}

where $S$ denotes the spin magnitude.

By computing all nine combinations of $\alpha, \beta \in \{x, y, z\}$, the full exchange tensor $\mathbf{J}_{ij}^{(1)}$ can be constructed, requiring a total of $4 \times 9 = 36$ calculations. This underscores the necessity of an automated workflow as provided by AMaRaNTA.

In the limit of perfectly isolated spin pairs -- achieved using sufficiently large supercells \cite{Menichetti_2019} -- and assuming that the Heisenberg Hamiltonian~\eqref{eq.spinH} captures the low-energy physics of the system, the four-state approach ensures that all extraneous contributions to the Hamiltonian cancel exactly. This guarantees that the extracted exchange parameters corresponds solely to the intended spin pair.

\vspace{0.35cm}
\paragraph*{Second- and third-nearest-neighbour scalar exchange parameters} 
The scalar exchange parameters $J^{(2)}_{ij}$ and $J^{(3)}_{ij}$, corresponding to pairs of second- and third-nearest-neighbour magnetic atoms $(i, j)$, are obtained using collinear DFT calculations without SOC, where spins are constrained to align along a fixed quantisation axis (in our case, the $z$-axis). Similarly to above, four distinct spin configurations are considered, in which the spins of atoms $i$ and $j$ are either parallel (i.e., up) or antiparallel (i.e., down) to the quantisation axis, and all other spins in the system are fixed in a collinear reference configuration. This yields four total energies: $E_{ij,\uparrow\uparrow}$, $E_{ij,\uparrow\downarrow}$, $E_{ij,\downarrow\uparrow}$, and $E_{ij,\downarrow\downarrow}$. The scalar exchange parameter is then given by

\begin{equation}
\label{4state_j2_j3}
J_{ij}^{(n=2,3)} = \frac{E_{ij,\uparrow\uparrow}-E_{ij,\uparrow\downarrow}-E_{ij,\downarrow\uparrow}+E_{ij,\downarrow\downarrow}}{4S^2}
\end{equation}

\vspace{0.35cm}
\paragraph*{Single-ion anisotropy} 
The SIA parameter is calculated using noncollinear DFT with SOC included. We consider four distinct spin configurations, in which the spin of atom $i$ is constrained to point along the $+x$, $-x$, $+z$, and $-z$ directions, while the spins of all other atoms are fixed along the $y$-axis. This yields the corresponding energies $E_{i,+x}$, $E_{i,-x}$, $E_{i,+z}$, and $E_{i,-z}$. 
The anisotropy parameter is then given by

\begin{equation}
\label{4state_sia}
A_i = \frac{
E_{i,+z} - E_{i,+x} - E_{i,-x} + E_{i,-z}
}{4S^2}
\end{equation}

\vspace{0.35cm}
In the actual implementation of AMaRaNTA, spin magnitudes $S$ are always normalised to unity. Accordingly, the reported exchange and anisotropy parameters correspond directly to the algebraic sums of energies defined in Eqs. (\ref{eq:j_tensor}), (\ref{4state_j2_j3}) and (\ref{4state_sia}), divided by 4.

\subsection*{Implementation}
\vspace{0.2cm}

\begin{figure*}[t] 
\includegraphics[width=0.70\textwidth]{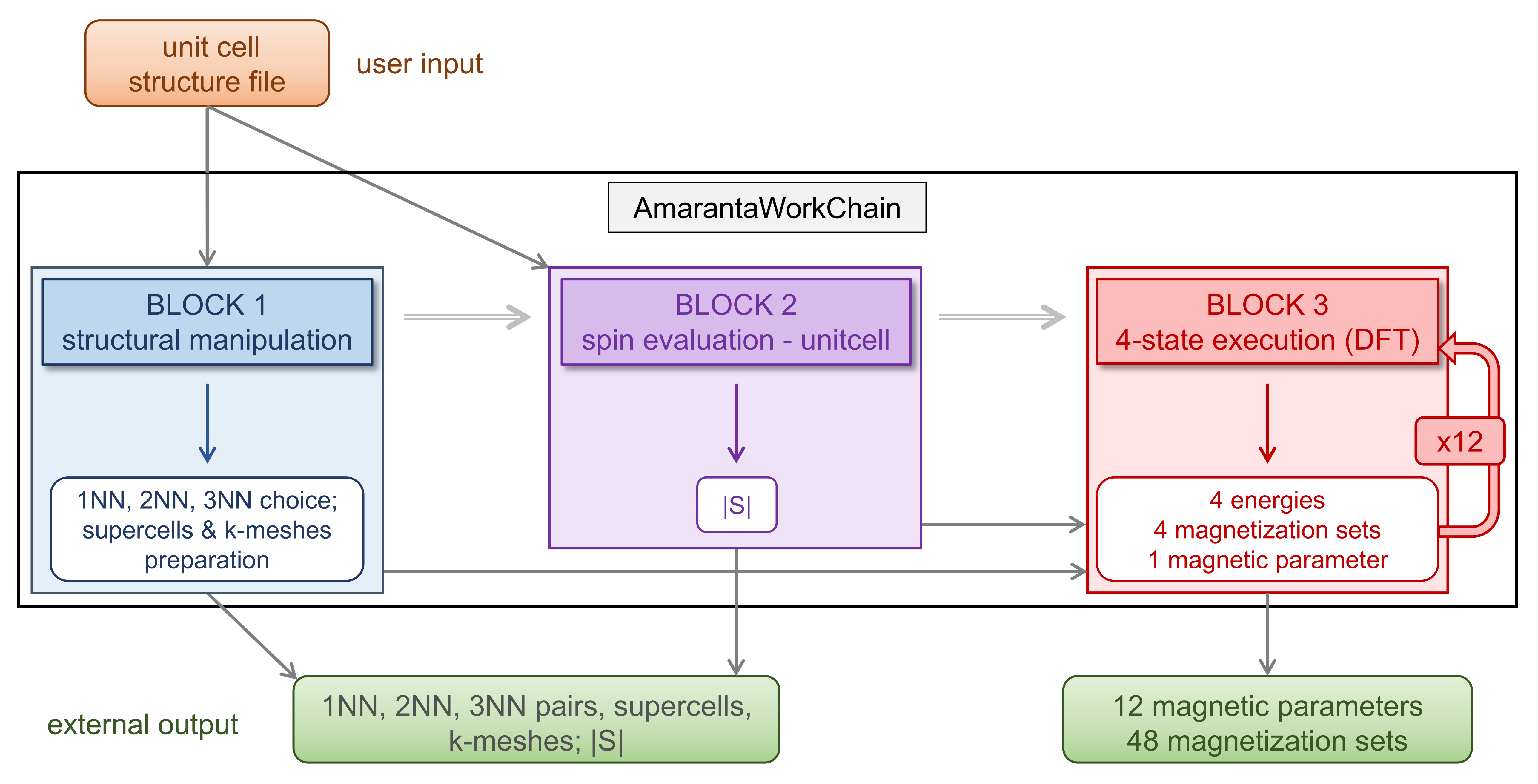}
\centering
\caption{Scheme of AMaRaNTA. Coloured boxes represent the three main blocks of the workflow, as discussed in the text. White boxes inside contain the partial output of each block. Thin arrows indicate direct input–output relations, thicker ones represent the procedural progression between blocks.}
\label{fig:map}
\end{figure*}

\subsubsection*{Workchain}

AMaRaNTA is implemented in Python as an AiiDA workchain \cite{Aiida_workchain_Uhrin}, specifically employing the AiiDA-VASP plugin (see https://aiida-vasp.readthedocs.io/en/latest/). It consists of three main blocks, as illustrated in Fig. \ref{fig:map}.

\vspace{0.35cm}
\paragraph*{Pair selection, supercell and reciprocal lattice construction}
The first block (blue in Fig. \ref{fig:map}) handles the identification of atomic pairs $(i,j)$ and the generation of appropriate supercells, using tools provided by the Atomic Simulation Environment (ASE) library \cite{Ase}. 

Starting from the user-supplied unit cell structure file, AMaRaNTA identifies three pairs $(i,j)$ of magnetic atoms corresponding to the nearest, second-nearest, and third-nearest-neighbour shells. To ensure all neighbours are accounted for, this step is performed in a 5$\times$5 supercell (employed at this stage only, then discarded afterwards). Atom $i$ is chosen once and for all and is thus the same in the three pairs. By default, it is selected as the magnetic atom closest to the centre of the supercell, which helps avoid complications arising from periodic boundary conditions when distant neighbours fall near the supercell edges. The partner atoms $j$ are chosen as the first three in the atomic list whose distances from $i$ place them within the nearest, second-nearest, or third-nearest-neighbour shell. Atoms whose distances from $i$ differ by less than 1\% are grouped within the same neighbour shell, thereby avoiding artefacts due to negligible atomic distance differences in the input structure.

Appropriate supercells are then constructed to ensure that each spin pair $(i,j)$ and its periodic replicas are separated by a distance of at least 10 \AA{}, a value we found sufficient for the systems studied in this work, that can, however, be easily increased if necessary. The initial guess for the supercell size is a 2$\times$2 of the input unit cell. From this starting point, two nested loops iterate over possible integer values $N$ and $M$, treated independently, to find the smallest $N\times M$ supercell that satisfies the distance criterion.

For a unit cell with lattice parameters $a$ and $b$, the $\mathbf{k}$-point mesh for the DFT calculations is defined as $\left( \text{round} \left( \frac{50}{a} \right),\text{round} \left( \frac{50}{b} \right), 1 \right)$ in reciprocal coordinates. When constructing the supercells, the $\mathbf{k}$-point meshes are rescaled inversely with respect to the corresponding supercell lattice vectors, ensuring a consistent sampling density across different cell sizes.

\vspace{0.35cm}
\paragraph*{Spin magnitude evaluation}

The second block (violet) performs a single DFT calculation for the unit cell, to determine the modulus of the magnetic moment $S$ for the magnetic species. This value is then rounded to the nearest integer and later used to initialise the magnetic configurations, according to the four-state prescriptions.

\vspace{0.35cm}
\paragraph*{Execution of four-state calculations}

The third and main block (red) orchestrates the DFT calculations required by the four-state method. For each of the twelve parameters in the spin model Hamiltonian, a child workchain is launched to perform the four constrained DFT calculations.
The resulting total energies are processed using Eqs. (\ref{eq:j_tensor}), (\ref{4state_j2_j3}) and (\ref{4state_sia}), yielding the exchange parameters and the SIA in meV. As anticipated, with spin magnitudes normalised to unity, the parameters are directly obtained as one quarter of the energy combinations defined therein. Finally, information on the converged magnetisation of all atoms - as obtained in the DFT calculations - is collected, to enable users to verify that the assigned spin directions have been respected (AMaRaNTA does not issue automated alerts if deviations occur). This also allows for an \emph{a posteriori} check of the assumption that all magnetic atoms indeed share the same spin magnitude $S$.

\subsubsection*{Inputs and outputs}

AMaRaNTA is managed and launched via a separate script, where users can specify the structure file and a set of pseudopotentials without modifying the workchain itself. This represents the only required input; all other settings are fixed internally to ensure consistency across the database. However, selected parameters — such as the 10-\AA{} pair-replica distance or the subset of exchange parameters to compute — can be modified through documented and accessible options. This design aims at ensuring uniformity across the database, while still offering enough flexibility for external users to adapt the high-throughput calculations to their specific needs.

The final output includes the coordinates of the three atomic pairs $(i,j)$, the constructed supercells, the calculated set of spin model Hamiltonian parameters, and the converged magnetisation data obtained from DFT for all atoms. Atomic coordinates, in particular, are  needed in that the form of the nearest-neighbour exchange tensor depends on the relative orientation of the selected pair. Therefore, providing them allows users to compare tensors obtained from differently oriented pairs via suitable coordinate transformations. All the above information is made available both as AiiDA data nodes and as user-friendly text files, which are also stored within the AiiDA database. Additionally, much of the content is printed to the standard output on-the-fly. 

\begin{figure*}[t] 
\includegraphics[width=\textwidth]{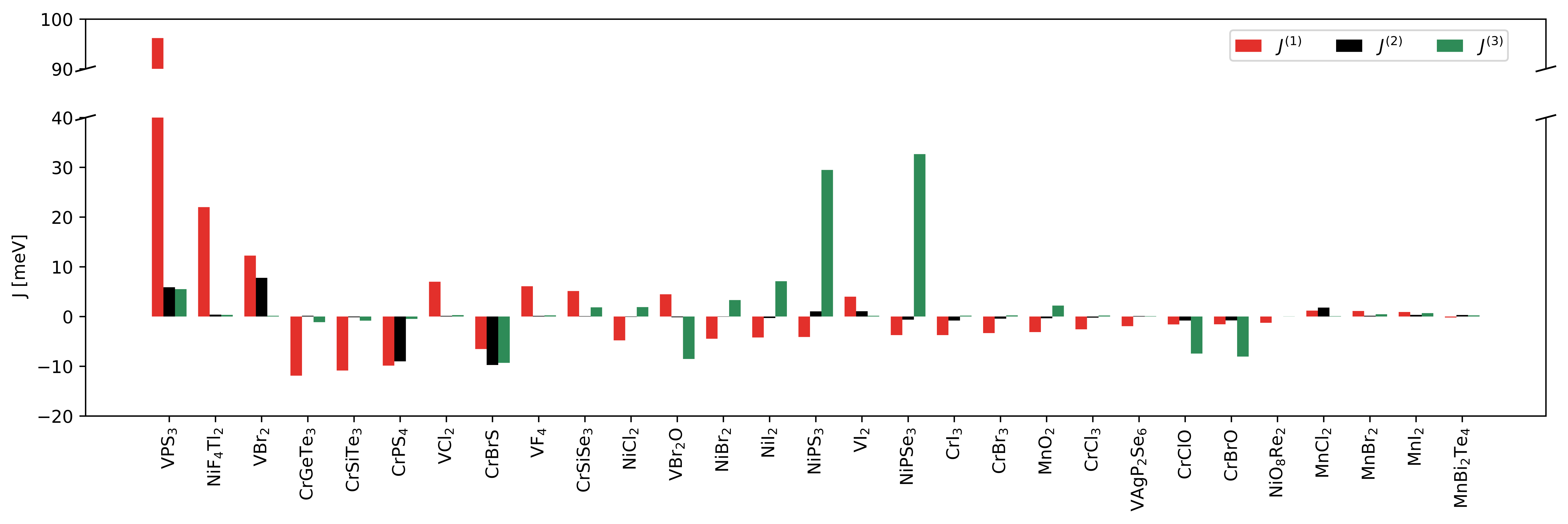}
\centering
\caption{Isotropic exchange parameters $J^{(1)}$, $J^{(2)}$ and $J^{(3)}$ for our dataset, shown in decreasing order of |$J^{(1)}$|.}
\label{fig:J123}
\end{figure*}

\vspace{0.25cm}
\subsection*{Application to a material dataset}
\vspace{0.2cm}

To illustrate the potential and performance of AMaRaNTA, we apply it to a dataset of 29 2D magnets, extracted from the Materials Cloud 2D Structure (MC2D) database \cite{MC2D_1,MC2D_2}. This dataset is limited to insulating compounds (primarily binary and ternary) with a calculated band gap larger than 0.1 eV at the Perdew-Burke-Ernzerhof (PBE) level, as reported in MC2D, and containing a single magnetic species — specifically, transition metals ranging from V to Ni — as mentioned above.

The results are analysed in terms of the magnitudes of the different magnetic parameters introduced in Eq.~(\ref{eq.spinH}), highlighting distinct trends across the various compounds and predicting whether their ground states are more likely to be collinear or noncollinear, based on the degree of magnetic frustration. The complete raw data of all calculations are provided in the Supplementary Material.

\vspace{-0.25cm}
\subsubsection*{Isotropic exchange}

The magnitude and sign of the isotropic exchange interactions — reported in Fig. \ref{fig:J123} for our dataset up to the third-nearest-neighbour level — are among the most fundamental descriptors of a magnetic material. As discussed above, the inclusion of these longer-range interactions marks a major advancement introduced by AMaRaNTA, distinguishing it from prior approaches.

The materials in our dataset can be broadly categorised into three distinct subsets, based on the relative magnitude of isotropic exchange: (i) those with a dominant nearest-neighbour interaction, (ii) those in which second- or third-nearest-neighbour interactions prevail, and (iii) those where interactions at different distances have comparable magnitudes, giving rise to competing effects. 

Hereafter, for convenience, we omit the atomic indices $i,j$ from the exchange parameters, wherever they are redundant. When comparing to external references, we adjust the reported exchange parameters when needed, so that they conform to our model Hamiltonian~\eqref{eq.spinH} and to our $S=1$ convention.

\vspace{0.35cm}
\paragraph*{Materials with dominant nearest-neighbour interaction} 

A significant subset of materials in our dataset exhibits a large isotropic exchange parameter between nearest-neighbour atoms and negligible interactions beyond. If this parameter is negative, such systems are expected to display a FM ground state, stable against thermal fluctuations in the presence of easy-axis anisotropy. 
Notable examples include the well-known compounds CrGeTe$_3$ and CrSiTe$_3$: our calculations reproduce the dominant negative $J^{(1)}$, consistent with collinear ferromagnetism, as reported in the literature \cite{CrGeTe_first, CrSiTe3_2015_Zhuang, Menichetti_2019}.

The most striking result in our dataset, however, is for VPS$_3$. We find a positive and exceptionally large $J^{(1)} = 96$ meV -- roughly 16–17 times larger than both $J^{(2)}$ and $J^{(3)}$ -- pointing to a robust AFM order. Comparing to previous theoretical studies, Torelli {\it et al.} \cite{Torelli} reported a $J^{(1)}$ value of 108 meV; Chittari {\it et al.} \cite{MPX3_2016_Chittari} found $J^{(1)} = 94.7$ meV, which was 6–8 times larger than $J^{(2)}$ and $J^{(3)}$; finally, Yang {\it et al.} \cite{VPS3_2025_Yang} predicted $J^{(1)} = 53.5$ meV, approximately 50 times larger than the higher-order terms. Despite some quantitative discrepancies -- possibly due to the different values of the Hubbard $U$ -- all the above predictions qualitatively agree in identifying the strongly AFM $J^{(1)}$ as the leading magnetic parameter. This conclusion is further supported by experimental evidence for a Néel-type AFM structure, consistent with the compound's honeycomb lattice \cite{VPS3_2023_Liu}. 

NiF$_4$Tl$_2$ is another example of a material with a dominant AFM $J^{(1)}$, for which no prior literature exists. This lack of data is likely due to thallium's toxicity, which limits experimental research.

Finally, our results also predict a leading AFM $J^{(1)}$ in the two families of dihalides MnX$_2$ and VX$_2$ (X = Cl, Br, I). However, unlike the previously discussed materials, these compounds crystallise in a structure where the magnetic ions form a triangular lattice. In such conditions, strong AFM interactions can give rise to geometric frustration, which may suppress simple collinear ordering and promote more complex magnetic states instead, as discussed, for example, in Ref.~\cite{2022_Riedl_Valentì}.

\vspace{0.35cm}
\paragraph*{Materials with dominant beyond-nearest-neighbour interactions}

Some materials in our dataset, notably NiPS$_3$ and NiPSe$_3$, are characterised by a very strong and AFM $J^{(3)}$ of about 30 meV, which dominates over a weak FM $J^{(1)}$ and a negligible $J^{(2)}$. This interaction pattern, when combined with the honeycomb lattice formed by the Ni atoms, leads to AFM spin textures -- either Néel or zigzag, depending on the specific values of the parameters -- as reported in many studies \cite{MPX3_2016_Chittari,MPX3_2019_Gu,MPX3_2021_Bazazzadeh,MPX3_2022_Basnet,MPX3_2024_Peng,MPX3_2025_Yang}. Tables \ref{tab:NiPS3} and \ref{tab:NiPSe3} summarise our results, compared to previous theoretical predictions and experimental measurements.

Most earlier results qualitatively agree with ours in identifying $J^{(3)}$ as the dominant magnetic interaction in these nickel compounds. Several works provide physical insight into this behaviour \cite{MPX3_2019_Gu,NiPS3_2022_Autieri,MPX3_2022_Basnet,MPX3_2025_Yang}, suggesting that $J^{(3)}$ arises from a Ni–X$_1$–X$_2$–Ni superexchange pathway entirely contained within a single sublattice. In contrast, the pathway responsible for $J^{(2)}$ involves two chalcogen atoms belonging to different sublattices, which weakens its effectiveness. 
Furthermore, the fact that \( J^{(3)} \) is even larger than \( J^{(1)} \) has been explained using a simple model considering direct exchanges only \cite{NiPS3_2022_Autieri}. Due to the trigonal crystal field, the Ni \( d \)-manifold splits into lower-energy even orbitals and higher-energy odd orbitals (with respect to the basal plane). Even orbitals are fully occupied because Ni\(^{2+}\) corresponds to a \( d^8 \) configuration: the only contribution to exchange comes therefore from the odd orbitals. For these, it has been shown \cite{NiPS3_2022_Autieri} that the third-nearest-neighbour couplings are indeed larger than the nearest-neighbour ones.

Interestingly, both NiPS$_3$ and NiPSe$_3$ are also reported in Ref. \cite{Torelli}, where only nearest-neighbour interactions are considered. That work assigns a large $J^{(1)}$ (around 30 meV) to both compounds. We speculate that this value may effectively incorporate longer-range contributions, particularly from $J^{(3)}$. This interpretation is further supported by results from a later study by the same group on NiPS$_3$ \cite{NiPS3_2021_Olsen}.

\begin{table}[h]
\centering
\caption{Exchange parameters for NiPS$_3$ from various studies.}
\begin{adjustbox}{max width=\columnwidth}
\begin{tabular}{lccc}
\hline
\textbf{Reference} & $J^{(1)}$ [meV] & $J^{(2)}$ [meV] & $J^{(3)}$ [meV] \\
\hline
This work      & $-4.1$ & $1.1$  & $29.5$ \\
Chittari et al~\cite{MPX3_2016_Chittari} (DFT)     & $-11.3$ & $-0.1$  & $36.0$ \\
Gu et al~\cite{MPX3_2019_Gu} (DFT)  & $-5.5$	& $-0.4$	&$45.5$ \\
Bazazzadeh et al~\cite{MPX3_2021_Bazazzadeh} (DFT)  & $-1.0$ & $-0.2$ & $3.9$  \\
Autieri et al~\cite{NiPS3_2022_Autieri} (DFT)  & $-3.3$ & $-0.2$ & $13.7$  \\
Olsen~\cite{NiPS3_2021_Olsen} (DFT) & $-2.6$ & $-0.3$ & $14.0$   \\
Basnet et al~\cite{MPX3_2022_Basnet} (DFT)         & $-3.5$ & $-0.3$ & $14.1$ \\
Peng et al~\cite{MPX3_2024_Peng} (DFT)              & $-2.5$ & /  & $9.5$   \\
Yang et al~\cite{MPX3_2025_Yang} (DFT)              & $-5.6$ & $0.7$  & $32.0$   \\
Lançon et al~\cite{NiPS3_2018_Lancon} (exp)             & $-3.8$ & $0.2$  & $13.8$ \\
\hline
\end{tabular}
\end{adjustbox}
\label{tab:NiPS3}
\end{table}

\begin{table}[h]
\centering
\caption{Exchange parameters for NiPSe$_3$ from various studies.}
\begin{adjustbox}{max width=\columnwidth}
\begin{tabular}{lccc}
\hline
\textbf{Reference} & $J^{(1)}$ [meV] & $J^{(2)}$ [meV] & $J^{(3)}$ [meV] \\
\hline
This work                                   & $-3.7$ & $-0.6$ & $32.7$ \\
Chittari et al~\cite{MPX3_2016_Chittari} (DFT)   & $-3.1$ & $1.0$  & $18.2$ \\
Gu et al~\cite{MPX3_2019_Gu} (DFT)  & $-9.3$	& $-0.1$	&$43.0$ \\
Bazazzadeh et al~\cite{MPX3_2021_Bazazzadeh} (DFT) & $-1.1$ & $-0.1$ & $4.0$  \\
Musari et al~\cite{NiPSe3_2022_Musari} (DFT) & $-11.3$ & $-0.3$ & $4.7$  \\
Basnet et al~\cite{MPX3_2022_Basnet} (DFT)         & $-4.5$ & $0.1$ & $16.1$ \\
Sun et al~\cite{NiPSe3_2023_Sun} (DFT) & $-7.9$ & $2.9$ & $1.5$  \\
Peng et al~\cite{MPX3_2024_Peng} (DFT)              & $-1.3$ & /  & $12.8$   \\
Yang et al~\cite{MPX3_2025_Yang} (DFT)              & $-7.6$ & $0.5$  & $40.4$   \\
\hline
\end{tabular}
\end{adjustbox}
\label{tab:NiPSe3}
\end{table}

\vspace{0.35cm}
\paragraph*{Materials with comparable exchange interactions} 

A considerable fraction of materials in our dataset features comparable exchange interactions involving multiple neighbours. 
Specifically, we identify 10 materials for which the ratio $J^{(2)}/J^{(1)}$ is as large as 25\%, and 14 for which $J^{(3)}/J^{(1)}$ reaches the same value.
Particularly interesting are those cases where $J^{(3)}$ (or $J^{(2)}$) and $J^{(1)}$ have opposite sign, as this is another typical situation that can give rise to magnetic frustration and noncollinear ground states.
Examples of such behaviour include NiX$_2$ (X = Cl, Br, I), NiPX$_3$ (X = S, Se), MnO$_2$, MnBi$_2$Te$_4$ and VBr$_2$O. We however emphasise that such constraints on the ratios are merely necessary conditions for noncollinear order to emerge: the geometry of the magnetic lattice plays an equally important role. For instance, VBr$_2$O, despite meeting the ratio-based criteria, should be excluded from the list of potentially frustrated systems, as the pattern given by an AFM $J^{(1)}$, a negligible $J^{(2)}$ and a FM $J^{(3)}$ on a square lattice actually precludes any kind of frustration.

\begin{figure}[h!] 
\includegraphics[width=\columnwidth]{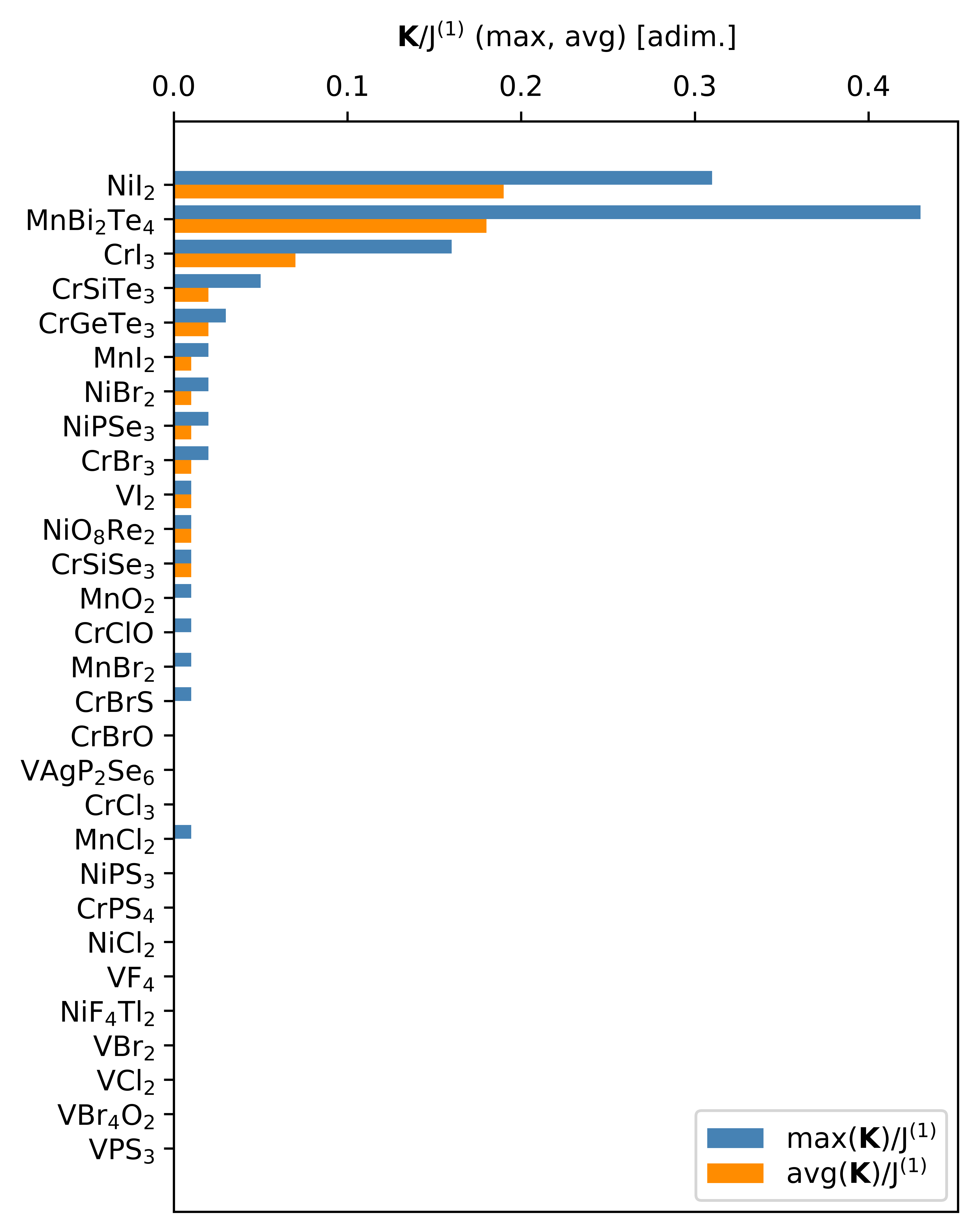}
\centering
\caption{Ratio between maximum (respectively, average) element of $\mathbf{\underline{\underline{K}}}$ and $J^{(1)}$.}
\label{fig:ST}
\end{figure}

\begin{figure}[h!] 
\includegraphics[width=\columnwidth]{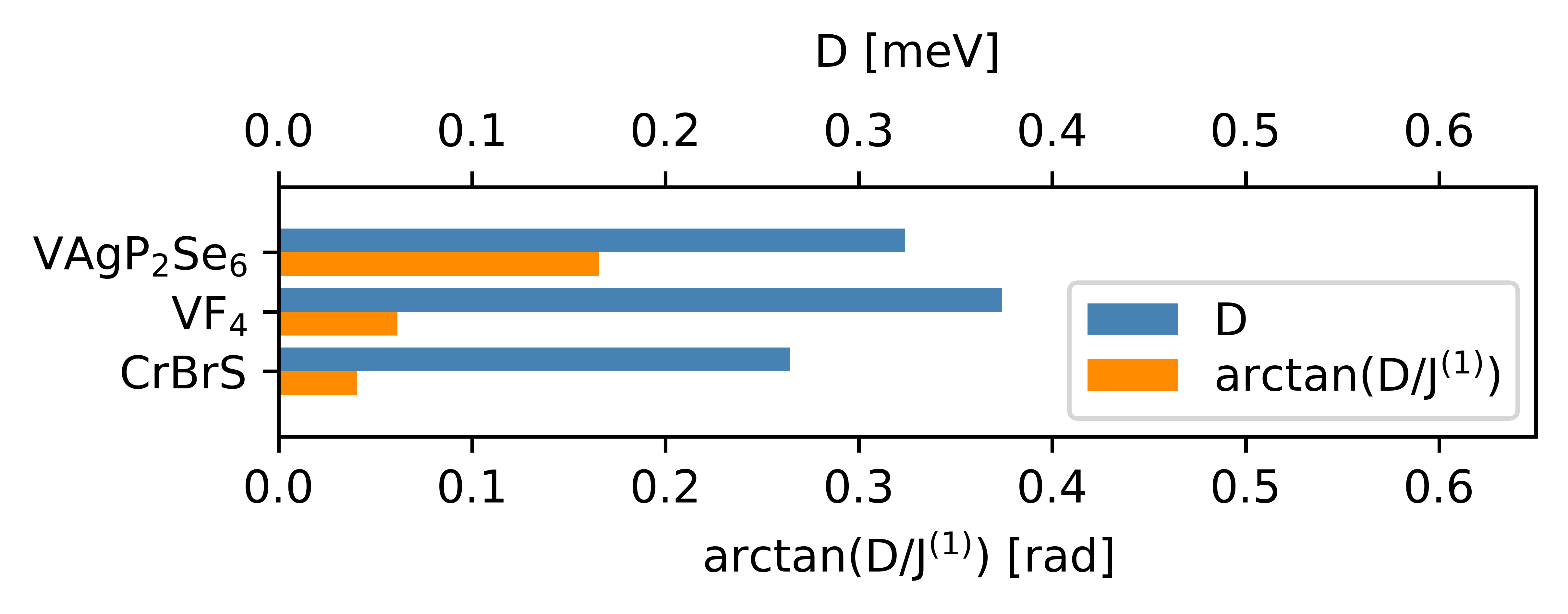}
\centering
\caption{Modulus of the Dzyaloshinskii–Moriya vector $D = |\mathbf{{D}}|$ and arctan(${D}/J^{(1)}$), shown for the only three materials with a non-vanishing antisymmetric exchange.}
\label{fig:AS}
\end{figure}

\subsubsection*{Magnetic anisotropies}
Magnetic anisotropy breaks the otherwise rotational symmetry of the Heisenberg Hamiltonian. It consists of two contributions: the SIA, denoted as 
$A$ in Eq.~(\ref{eq.spinH}), and the two-ion anisotropic exchange, captured by the symmetric Kitaev--like matrix $\mathbf{\underline{\underline{K}}}$ and the antisymmetric DMI matrix $\mathbf{\underline{\underline{D}}}$ in Eq.~(\ref{eq.J1decomp}). The ultimate preferential spin direction is decided by the interplay of the two contributions, which should therefore be treated on equal footing (see, e.g., Ref. \cite{Xiang_CrI3} for a practical prescription). Moreover, two-ion anisotropic exchange can itself be a driving force behind exotic, noncollinear spin textures \cite{Amoroso2020}, along with the effects of lattice geometry and competing isotropic exchange interactions discussed above.

The accurate evaluation of anisotropies -- in particular, the two-ion contribution -- beyond the approximations used in previous works constitutes the second crucial improvement provided by our approach. 

\vspace{0.35cm}
\paragraph*{Two-ion anisotropic exchange}
Fig. \ref{fig:ST} presents both the maximum and the average element of $\mathbf{\underline{\underline{K}}}$ (in absolute value) normalised by the isotropic term $J^{(1)}$. As expected, the strongest anisotropies are found in materials containing heavy elements such as tellurium and iodine, where SOC is more pronounced. We identify three materials where the average Kitaev--like interaction reaches at least 10\% of $J^{(1)}$. For two of them, the Kitaev--like magnetism is already well known in the literature: in CrI$_3$, it enhances the easy-axis anisotropy in cooperation with the SIA \cite{Xiang_CrI3}, whereas in NiI$_2$ its interplay with the strong $J^{(3)}$ induces the skyrmionic texture predicted in \cite{Amoroso2020}. The remaining one, MnBi$_2$Te$_4$, gained a certain attention some years ago due to its topological properties \cite{MnBi2⁢Te4_2019_Zhang}. The monolayer ground state is widely reported to be FM with out-of-plane anisotropy, in agreement with our prediction \cite{MnBi2⁢Te4_2019_Li, MnBi2⁢Te4_2019_Otrokov, MnBi2⁢Te4_2020_BingLi}. In contrast, it is listed in the database of Ref.~\cite{Torelli} as weakly AFM, but switching to a FM state in DFT+U calculations for a Hubbard interaction $U>2$ eV. Notably, no previous work has investigated the two-ion anisotropic terms in this system, except for \cite{MnBi2⁢Te4_2019_Li}. However, that study found no significant Kitaev--like contribution, at variance with our results. A deeper investigation of this material is therefore warranted in future studies.

Fig. \ref{fig:AS} shows both the magnitude $D$ of the Dzyaloshinskii–Moriya vector and the angle arctan(${D}/J^{(1)}$),   
which governs the degree of noncollinearity between spins. DMI can arise only when the bond centre lacks inversion symmetry, imposing a strong structural constraint. Accordingly, non-vanishing DMI is found in only three materials throughout our dataset. For CrBrS, a comprehensive study of the DMI -- also extended to beyond nearest-neighbours -- found it to be crucial for the formation of a topological bimeron texture \cite{CrSBr_2025_Baishun}. On the other hand, VF$_4$ has recently been highlighted for its altermagnetic behaviour \cite{Sodequist2024May} and VAgP$_2$Se$_6$ was treated in detail in \cite{VAgP2Se6_2023_Olsen} as a paradigmatic case of a polar 2D magnet. To the best of our knowledge, though, no DMI analysis has been reported to date for any of the two.

\vspace{0.35cm}
\paragraph*{Single-ion anisotropy}

\begin{figure}[h] 
\includegraphics[width=\columnwidth]{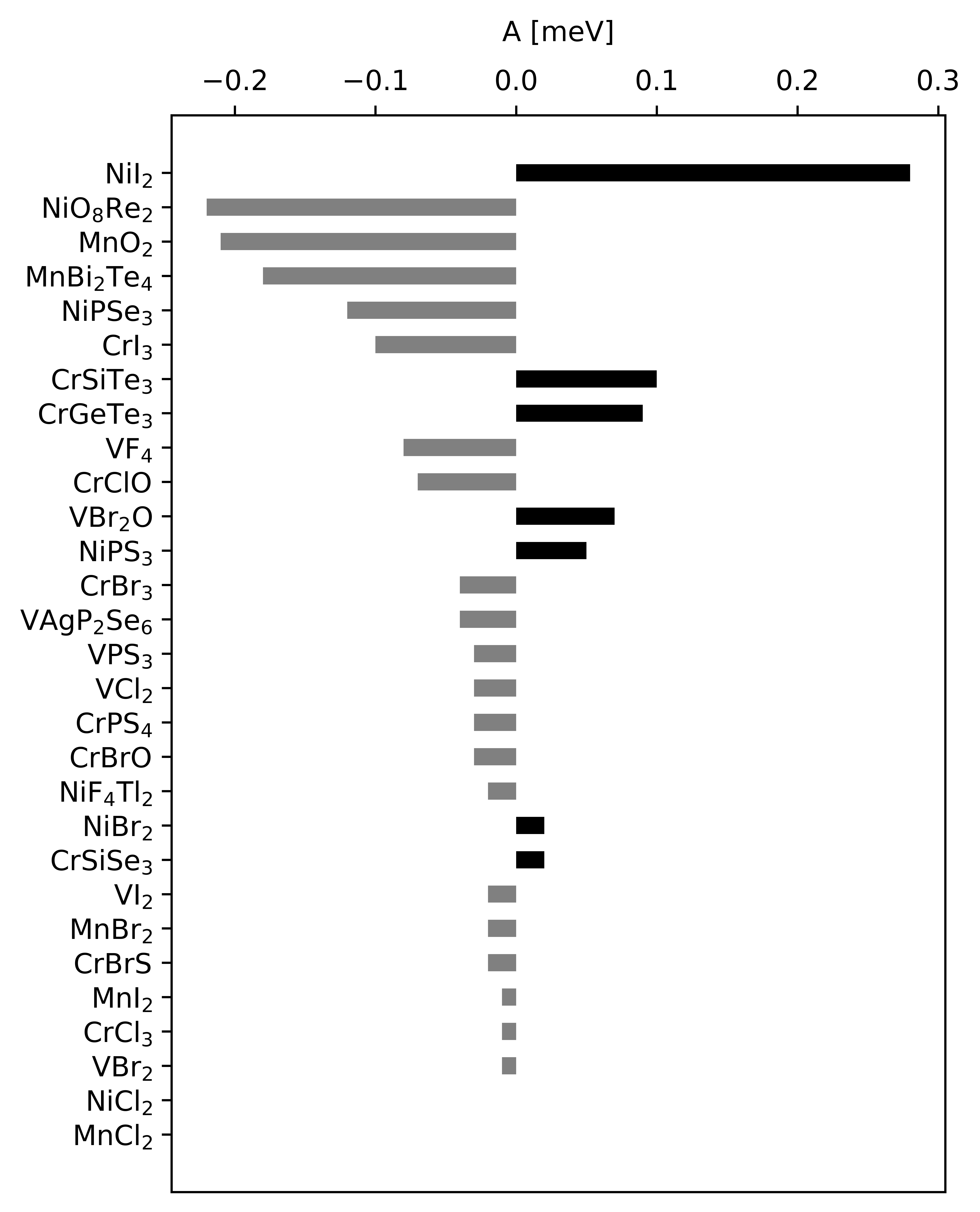}
\centering
\caption{SIA parameter $A$.}
\label{fig:sia}
\end{figure}

Finally, in Fig. \ref{fig:sia} we report the values of the SIA. We identify eight materials exhibiting a significant such value ($A \gtrsim$ 0.1 meV). Three of them -- NiI$_2$, CrSiTe$_3$ and CrGeTe$_3$ -- show $A$ > 0, pointing to an easy-plane tendency. The remaining five -- NiO$_8$Re$_2$, MnO$_2$,  MnBi$_2$Te$_4$, NiPSe$_3$ and CrI$_3$ -- exhibit $A$ < 0, suggesting an easy-axis tendency instead. All predictions are in agreement with existing literature \cite{Xiang_CrI3,Amoroso2020,MPX3_2022_Basnet,MnBi2⁢Te4_2019_Li,Haastrup2018Sep,Gjerding2021Jul}.

We stress again, however, that the sign of the SIA alone does not fully determine the overall magnetic anisotropy. For instance, CrGeTe$_3$ and CrSiTe$_3$, which would be expected to favour easy-plane configurations based on their positive $A$, actually do not, owing to the counteracting effect of the Kitaev--like term. In CrGeTe$_3$, the single-ion and Kitaev--like contributions essentially cancel out, resulting in an almost isotropic magnetic ground state. In CrSiTe$_3$, the Kitaev--like term outweighs the SIA, ultimately producing an easy-axis texture \cite{Xiang_CrI3}.

\section*{DISCUSSION}

We outlined the features of AMaRaNTA, an open-source computational workflow designed to automate DFT calculations of exchange parameters in 2D magnets. The full nearest-neighbour exchange tensor, scalar exchange parameters for second- and third-nearest-neighbours, and the single-ion anisotropy were included in the underlying spin model, providing a comprehensive and robust description of the minimal interactions required to stabilise the complex magnetic states observed in 2D materials.

To validate its functionality, we tested AMaRaNTA on a compact dataset of 2D magnets, consistently recovering the expected results for well-known materials, while also providing the added value of a uniform description across the dataset -- namely, with the same set of parameters. In selected cases, our calculations even suggested previously unreported exchange phenomena, demonstrating the potential of our package for applications in material discovery.

In this context, AMaRaNTA contributes to the growing ecosystem of computational tools to automate the calculation of the magnetic parameters of 2D magnets. This, in turn, enables systematic comparison of four-state results with alternative approaches, such as energy-mapping based on global magnetic configurations, as implemented in the recently released OstravaJ \cite{OstravaJ}, or the LKAG method used in the TB2J code \cite{TB2J} as well as in another recent package \cite{PhysRevB.99.224412, PhysRevB.108.214418}.

Beyond the materials considered in this work, AMaRaNTA can already be employed in its current version for larger-scale studies, such as investigations of rare-earth–based compounds \cite{rare_earth_magnets} or of the effect of strain on exchange interactions, which can be substantial, as shown for NiI$_2$ \cite{Liu_2025_NiI2_strain}. In this case, the workflow’s automation infrastructure enables systematic exploration of the strain parameter space with minimal manual effort.

In terms of future developments, AMaRaNTA can be extended in several directions with relatively modest additional endeavour. 
First, from a methodological viewpoint, the spin model could be generalised to capture more complex behaviours.
Examples include beyond-nearest-neighbour anisotropic exchange \cite{Orozovic_2025} or isotropic exchange beyond the third-nearest-neighbour level \cite{J4_CrTe2} as well as refining the single-ion anisotropy by lifting the current assumption of in-plane isotropy -- an improvement particularly relevant for systems with spatially anisotropic lattices. At the same time, the framework could also be expanded to handle symmetry-inequivalent magnetic sites, which requires determining more than a single set of parameters and would significantly broaden the class of materials that can be addressed. Even more ambitiously, the four-state protocol underpinning AMaRaNTA could be extended beyond the evaluation of exchange and anisotropy parameters, for instance to compute the coupling matrix between spin and polarisation in the context of spin-induced ferroelectricity, which relies on a similar computational machinery \cite{Xiang2011Oct,Edstrom2025Mar}.

Second, from the DFT perspective, AMaRaNTA could be extended to adopt material-specific, first-principles values for the Hubbard $U$ parameter \cite{Self_cons_U_Cococcioni, Self_cons_U_cRPA}, which is kept fixed in the current implementation (as detailed in the Methods section). Although qualitative trends are generally robust against moderate variations in $U$, the choice of this parameter can affect the results for materials where the computed exchange interactions are small, as exemplified by MnBi$_2$Te$_4$. In such cases, computing $U$ from first principles would remove much of the ambiguity associated with an empirical or arbitrary choice. Moreover, AMaRaNTA is not intrinsically tied to the DFT+$U$ framework and could equally be combined with more advanced exchange–correlation schemes, including hybrid and self-interaction–corrected functionals, which are known to impact exchange interactions \cite{Archer2011Sep}.

Finally, on the software side, two natural extensions would involve making AMaRaNTA compatible with DFT codes beyond VASP, provided they support constrained-magnetisation calculations, and developing a streamlined interface to Monte Carlo or atomistic spin-dynamics simulation packages. This would pave the way toward a fully automated pipeline for the systematic, high-throughput exploration of complex magnetic phenomena in 2D materials.

\section*{METHODS}

\paragraph*{DFT computational details}
DFT calculations are carried out using VASP \cite{VASP1,Vasp2} (version 6.4.0). The Local Spin-Density Approximation (LSDA) is employed for the exchange-correlation functional, which avoids the issue of symmetry breaking known to occur in noncollinear calculations with the Generalised Gradient Approximation (GGA) \cite{GGAnoncoll}. For magnetic species, the so-called ``\_pv''-type pseudopotentials are used, treating semicore $p$ states as valence electrons \cite{pseudo1,pseudo2}. Electron correlation is treated at the LDA$+U$ level, following the Liechtenstein scheme \cite{Liechtenstein}, with $U = 2.8$ eV and $J = 0.8$ eV. This set of parameters was applied uniformly to all magnetic species in the studied compounds, to ensure consistency across our dataset. The plane-wave kinetic energy cutoff is set to 1.3 times the largest ENMAX value among all elements in each particular system. Total energies are converged to within $10^{-6}$ eV. All symmetry operations are explicitly disabled to ensure that every element of the nearest-neighbour exchange tensor is computed independently.

\vspace{0.35cm}
\paragraph*{Precision of exchange parameters}
We estimate the precision of our results to be on the order of 0.1 meV, even though we report an additional significant digit. Extensive testing indicates that this uncertainty is primarily determined by the choice of supercell size. Secondary factors, contributing 0.01–0.04 meV, include the energy cutoff, $\mathbf{k}$-point mesh density, and small deviations arising from unfulfilled symmetry constraints.

In this analysis, only parameters requiring convergence are treated as sources of error. Categorical choices, such as the pseudopotential family or type, are not included, as they do not affect the internal consistency of the database. Consequently, users attempting to reproduce our results with different choices may observe larger deviations. Additional discrepancies, estimated at roughly 0.1–0.15 meV, can also arise from differences in VASP versions or variations in hardware and high-performance computing environments.

\section*{ACKNOWLEDGEMENTS}

The authors acknowledge (in alphabetical order) Jakob Baumsteiger, Guido Menichetti, Lei Qiao, Alessandro Stroppa, and Cesare Tresca for useful discussions.
The necessary structure files for the dataset generation were provided as a courtesy by Davide Campi.

The work was funded by the European Union - NextGenerationEU, through the ICSC-Centro Nazionale di Ricerca in High-Performance Computing, Big Data and Quantum Computing (Grant No. CN00000013, CUP J93C22000540006, PNRR Investimento M4.C2.1.4). A.M. and M.G. acknowledge partial support from the PRIN Project “Simultaneous electrical control of spin
and valley polarisation in van der Waals magnetic materials” (SECSY–CUP Grant Nos. E53D23001700006 and J53D23001400001, PNRR
Investimento M4.C2.1.1), which is funded by the European Union - NextGenerationEU. S.P. acknowledges partial support from the PRIN Project “SORBET—Spin-orbit effects in two-dimensional magnets” (IT-MIUR Grant No. 2022ZY8HJY) within NextGenerationEU. S.P. and G.C. acknowledge partial support from the National Quantum Science and Technology Institute (NQSTI) within PNRR MUR Project No. PE0000023-NQSTI  within NextGenerationEU. The views and opinions expressed are solely those of the authors and do not necessarily reflect those of the European Union, nor can the European Union be held responsible for them. 

S.S. acknowledges financial support from the Vin\v{c}a Institute, provided by the Ministry of Science, Technological Development, and Innovation of the Republic of Serbia. S.S., P.B., A.D., G.C. and S.P. acknowledge additional funding from the Ministry of Foreign Affairs of Italy and the Ministry of Science, Technological Development, and Innovation of Serbia through the bilateral project ``Van der Waals Heterostructures for Altermagnetic Spintronics'', realised under the executive programme for scientific and technological cooperation between the two countries.

We acknowledge the CINECA award under the ISCRA initiative (project ISCRA-B HP10BA00W3), for the availability of high-performance computing resources and support.

\section*{DATA AVAILABILITY}
The dataset generated and analysed in the current study is available in the Supplementary Materials. 

\section*{CODE AVAILABILITY}
The underlying code for this study is not publicly available at this time, but will be released in the future. In the meantime, it can be shared with qualified researchers upon reasonable request.  

\section*{AUTHOR INFORMATION}

\paragraph*{Contributions}
F.O. software preparation, execution, supervision, data analysis, writing – original draft; A.D. execution, supervision, writing – original draft; L.V., C.F. software preparation; G.C. execution; P.B. supervision; A.M., M.G. conceptualisation, supervision; S.S. conceptualisation, software preparation; S.P. conceptualisation, supervision. All authors participated in the revision and editing of the manuscript and approved the final version.

\vspace{0.35cm}
\paragraph*{Corresponding author}
Correspondence to Federico Orlando (federico.orlando@polimi.it).

\section*{COMPETING INTERESTS}
All authors declare no financial or non-financial competing interests. 
\vspace{0.5cm}

\bibliography{main.bbl}

\end{document}